\documentclass[groupedaddress,aps,pra,superscriptaddress,showpacs,twocolumn,prl]{revtex4}%
\usepackage{epsfig,dsfont,amssymb,amsmath,amsthm,amsfonts,amsbsy,mathrsfs}
\usepackage{graphicx, color}
\usepackage{epstopdf}
\usepackage{mathdots}
\usepackage{float}
\usepackage{xcolor}
\begin{document}

\title{Informationally complete measures of quantum entanglement}

\author{Zhi-Xiang Jin}
\affiliation{School of Physics, University of Chinese Academy of Sciences, Yuquan Road 19A, Beijing 100049, China}
\affiliation{Max-Planck-Institute for Mathematics in the Sciences, Leipzig 04103, Germany}
\author{Shao-Ming Fei}
\thanks{Corresponding author: feishm@cnu.edu.cn}
\affiliation{Max-Planck-Institute for Mathematics in the Sciences, Leipzig 04103, Germany}
\affiliation{School of Mathematical Sciences, Capital Normal University,  Beijing 100048,  China}
\author{Xianqing Li-Jost}
\affiliation{Max-Planck-Institute for Mathematics in the Sciences, Leipzig 04103, Germany}
\author{Cong-Feng Qiao}
\thanks{Corresponding author: qiaocf@ucas.ac.cn}
\affiliation{School of Physics, University of Chinese Academy of Sciences, Yuquan Road 19A, Beijing 100049, China}
\affiliation{CAS Center for Excellence in Particle Physics, Beijing 100049, China\\ \vspace{7pt}}

\begin{abstract}
Although quantum entanglement has already been verified experimentally and applied in quantum computing, quantum sensing and quantum networks, most of the existing measures cannot characterize the entanglement faithfully. In this work, by exploiting the Schmidt decomposition of a bipartite state $|\psi\rangle_{AB}$, we first establish a one-to-one correspondence relation between the characteristic polynomial of the reduced state $\rho_A$ and the polynomials its trace. Then we introduce a family of entanglement measures which are given by
the complete eigenvalues of the reduced density matrices of the system.
Specific measures called informationally complete entanglement measures (ICEMs) are presented to illustrate the advantages. It is demonstrated that such ICEMs can characterize finer and distinguish better the entanglement than existing well-known entanglement measures. They also give rise to criteria of state transformations under local operation and classical communication. Moreover, we show that the ICEMs can be efficiently estimated on a quantum computer. The fully separability, entanglement and genuine multipartite entanglement can detected faithfully on quantum devices.
\end{abstract}
\maketitle

{\it Introduction}--Quantum entanglement is the striking feature of quantum states which exhibits correlations that cannot be accounted for classically. It is the fundamental resource and plays an important role in many quantum information processing such as quantum communications \cite{bch,jb,rc,ng}, quantum cryptography \cite{jmr,ml,ake,ngg} and quantum computing \cite{ae,nie,dat}. Quantifying the entanglement has become a very important issue for both theoretical and potentially practical reasons.

Various measures of entanglement have been proposed in recent years \cite{HPB,HPBO,JIV,CYSG,RPMK,con, wootters,wootters2,pmb}, such as concurrence \cite{con, mf,KSS,FMA}, negativity \cite{zkp,vidal}, entanglement of formation \cite{con, wootters,vgl}, relative entropy of entanglement \cite{pmb} and Renyi entropy entanglement \cite{vgl}.
In Ref. \cite{smm,fanh,gour,bms}, the authors presented a family of concurrence monotones by means of the polynomials of the Schmidt coefficients, in order to show that these measures can characterize more information related to entanglement. Recently, in Ref. \cite{jacob} the authors provided a family of multipartite entanglement measures called concentratable entanglements. In \cite{lmx}, the the authors proposed a new parameterized bipartite entanglement monotone named $q$-concurrence by general Tsallis entropy.
These quantum entanglement measures make further progresses on the characterization of quantum entanglement. In fact, these measures are given by part of the Schmidt coefficients of a given pure entangled state. In fact, the characterization of the entanglement can be improved by taking into account of all the Schmidt coefficients of a pure entangled state.

In this paper, we first give a one-to-one correspondence between the coefficients of the characteristic polynomial of a quantum state $\rho$ and the traces of its powers. For a pure state $|\psi\rangle$ the trace of the power of its reduced state quantifies completely the correlations as all the Schmidt coefficients of $|\psi\rangle$ have been taken into account. Then, we introduce a family of quantities called informationally complete entanglement measures (ICEMs) to characterize the entanglement of arbitrary $n$-qudit pure states. For any mixed state $\rho$, we define the ICEMs via the average on that of pure states in the pure state decomposition of $\rho$, minimized over all possible pure state decompositions of $\rho$. We prove that each ICEM does not increase under local quantum operations assisted with classical communications on average, namely, these ICEMs are entanglement monotones.
We present examples to show the superiorities of our ICEM in identifying quantum entanglement by comparing with other existing measures. Moreover, we show that the ICEM is related to the transformation between states under local operation and classical communication (LOCC). Finally, we discuss how to efficiently estimate the ICEMs on a quantum computer.

{\it Informationally complete entanglement measures (ICEMs)}--Let $|\psi\rangle_{AB}$ be a pure state of a composite system $A$ and $B$ with individual dimension $d$. In suitable local bases a bipartite pure state $\psi_{AB}$ has the Schmidt decomposition form, $|\psi\rangle_{AB}=\sum_{i=1}^d \sqrt{\lambda_i}|ii\rangle$, where $\lambda_i$ are the Schmidt coefficients satisfying $\sum_i^d \lambda_i=1$.
Denote $\rho_A$ and $\rho_B$ the reduced density matrices of $\rho_{AB}=|\psi\rangle_{AB}\langle \psi|$ with respect to the subsystems $A$ and $B$, respectively. $\rho_A$ and $\rho_B$ have the same eigenvalues given by the Schmidt coefficients $\lambda_i$. Many important properties of a quantum state are contained in the eigenvalues of the reduced density matrices. A reduced state $\rho_A$ has the characteristic polynomial,
\begin{eqnarray}\label{eigen}
F(\lambda)&=&\Pi_{i=1}^d(\lambda-\lambda_i)\nonumber\\
&=&a_0\lambda^d-a_1\lambda^{d-1}+a_2\lambda^{d-2}\nonumber\\
&&+\cdots+(-1)^{d-1}a_{d-1}\lambda+(-1)^{d}a_d,
\end{eqnarray}
where $a_0=1$, $a_k=\sum_{\{i_j\}\subseteq\mathcal{S}}\Pi_{j=1}^k\lambda_{i_j}$ for $k=1,2,...,d$, $\{i_j\}$ is a subset of $\mathcal{S}$ ($\mathcal{S}=\{1,2,...,d\}$) with $k$ elements. Thus, the entanglement properties of a state $|\psi\rangle_{AB}$ are encoded in all the coefficients $a_k$ of the characteristic polynomial. For instance, the generalized concurrence \cite{pr,mf},
\begin{eqnarray}\label{concur}
C(|\psi\rangle_{AB})=\sqrt{{\frac{d}{d-1}\left[1-\mathrm{Tr}(\rho_A^2)\right]}},
\end{eqnarray}
is given by the coefficient $a_2$ as $1-\mathrm{Tr}(\rho_A^2)=2a_2$.

Denote $\mathcal{A}(\rho_A)=\{a_1, a_2,...,a_d\}$, with $a_1, a_2,...,a_d$ the coefficients of the characteristic polynomial of $\rho_A$ as given in Eq. (\ref{eigen}), and $\mathcal{T}(\rho_A)=\{\mathrm{Tr}(\rho_A),\mathrm{Tr}(\rho^2_A),...,\mathrm{Tr}(\rho^d_A)\}$. The number of elements in the sets $\mathcal{A}(\rho_A)$ and $\mathcal{T}(\rho_A)$ are related to the Schmidt rank of $|\psi\rangle_{AB}$.
We first present a one-to-one correspondence between the set $\mathcal{A}(\rho_A)$ and $\mathcal{T}(\rho_A)$, see proof in the Supplemental Material I.

{\it Lemma}--Given an bipartite pure quantum state $|\psi\rangle_{AB}$, the sets of $\mathcal{A}(\rho_A)$ and $\mathcal{T}(\rho_A)$ have the following relations,
\begin{eqnarray}\label{th}
a_{k+1}&=&\frac{1}{k+1}\sum_{l=0}^k (-1)^l a_{k-l}\mathrm{Tr}(\rho^{l+1}_A),
\end{eqnarray}
where $a_0=1$, $k=0,1,...,d-1$.

Given a pure quantum state $|\psi\rangle_{AB}$ with Schmidt rank $d=r+1$ (the number of nonzero Schmidt coefficients), by using step-by-step the iterations of Eq. (\ref{th}) one can get each element $a_i$ of $\mathcal{A}(\rho_A)$ as the linear functions of the elements $\rho^i_A$ in $\mathcal{T}(\rho_A)$. Furthermore, from the relation
$a_k=\sum_{\{i_j\}\subseteq\mathcal{S}}\Pi_{j=1}^k\lambda_{i_j}$, $k=1,2,...,d$,
one can get the eigenvalues of $\rho_A$. Namely, the quantum pure state $|\psi\rangle_{AB}$ is fully characterized by $\mathcal{T}(\rho_A)$. We can obtain the following
informationally complete entanglement measure, see proof in the Supplemental Material II.

{\it Theorem 1.}--For bipartite state $|\psi\rangle_{AB}$ with Schmidt rank $R(|\psi\rangle_{AB})=r+1$, the ICEM is defined by
\begin{eqnarray}\label{def}
\mathcal{C}(|\psi\rangle_{AB})=1-\frac{1}{2^r}\sum_{i=0}^rP_r^i\mathrm{Tr}(\rho^{i+1}_A),
\end{eqnarray}
is a well defined measure of quantum entanglement, where $P_r^i=\frac{r!}{(r-i)!}$.

For mixed state $\rho_{AB}$, the ICEM is defined by the convex roof extension, $\mathcal{C}(\rho_{AB})=\min_{\{p_j,|\psi_j\rangle_{AB}\}}\sum_jp_j\mathcal{C}(|\psi_j\rangle_{AB})$ with the minimization taking over all possible pure state decompositions  $\{p_j,|\psi_j\rangle_{AB} \}$ of $\rho_{AB}$.
In fact, one can use the convex combination of any elements in set $\mathcal{T}(\rho_A)$ to define an entanglement measure. The measure Eq. (\ref{def})
contains all the elements in the set $\mathcal{T}(\rho_A)$.
According to Eq. (\ref{def}), when $R(|\psi\rangle_{AB})=2$ ($r=1$), we have $\mathcal{C}(|\psi\rangle_{AB})=\frac{1}{2}(1-\mathrm{Tr}(\rho_A^2))$, which coincides with the traditional entanglement measure concurrence (\ref{concur}). It implies that for fewer parameterized states, their entanglement can be characterized in a simpler way.
However, the concurrence does not capture the complete information of entanglement for higher ranked states.

{\it Advantages and applications}--To show the advantage of the ICEM, let us make a comparison with some other entanglement measures.
In Ref. \cite{jacob}, the authors introduce a family of quantities called concentratable entanglements $\widetilde{C}(|\psi\rangle_{AB})$ by using the convex of local purities $\mathrm{Tr}(\rho_A^2)$, $\widetilde{C}(|\psi\rangle_{AB})=1-\frac{1}{2^s}\sum_{\{A'\}}\mathrm{Tr}\rho_{A'}^2$, where $\{A'\}$ is a subset of $\{A\}$, $s$ is the cardinality of the set $\{A\}$, and the summation goes over all possible subset $\{A'\}$.
Consider the following rank-3 states: $|\phi_1\rangle_{AB}=\sqrt{\frac{1}{2}}|00\rangle+\sqrt{\frac{1}{3}}|11\rangle+\sqrt{\frac{1}{6}}|22\rangle$
and $|\phi_2\rangle_{AB}=\sqrt{\beta_1}|00\rangle+\sqrt{\beta_2}|11\rangle+\sqrt{1-\beta_1-\beta_2}|22\rangle$. By calculation (see proof in Supplemental Material III), we have that $C(|\phi_1\rangle_{AB})=C(|\phi_2\rangle_{AB})$ for the concurrence defined in Eq. (\ref{concur}) and $\widetilde{C}(|\phi_1\rangle_{AB})=\widetilde{C}(|\phi_2\rangle_{AB})$ for the measure given in Ref. \cite{jacob} for all the states $|\phi_2\rangle_{AB}$ in the ellipse in Fig. 1.
\begin{figure}
\centering
\includegraphics[width=7cm]{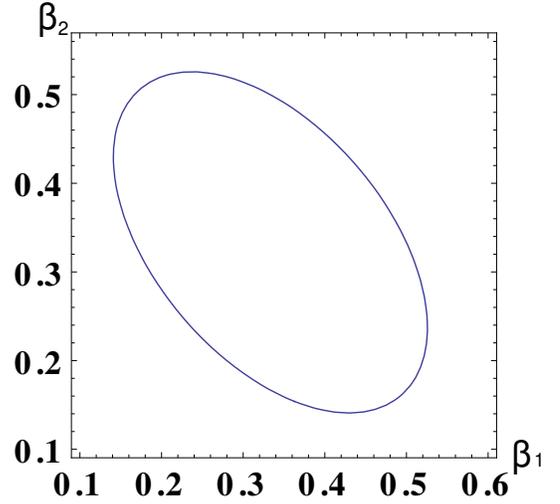}\\
\caption{ The horizontal and vertical axes represent $\beta_1$ and $\beta_2$, respectively. The points on the ellipse represent the states satisfying $C(|\phi_1\rangle_{AB})=C(|\phi_2\rangle_{AB})$ and $\widetilde{C}(|\phi_1\rangle_{AB})=\widetilde{C}(|\phi_2\rangle_{AB})$.}\label{2}
\end{figure}

However, by using the ICEM in Eq. (\ref{def}), one can verify that $\mathcal{C}(|\phi_1\rangle_{AB})$ is not equal to $\mathcal{C}(|\phi_2\rangle_{AB})$ for all the states $|\phi_2\rangle_{AB}$ in the ellipse except for three states given by the intersections of the solid red and the dot-dashed blue lines in Fig. 2.
These three intersections are given by $(\beta_1,\beta_2)=(1/2,1/3)$, $(1/3,1/2)$ and $(1/6,1/2)$, whose corresponding states are exactly $|\phi_1\rangle_{AB}$. Therefore,
our measure ICEM distinguishes the entanglements of $|\phi_1\rangle_{AB}$ and $|\phi_2\rangle_{AB}$ as long as $|\phi_1\rangle_{AB}$ is not equal to $|\phi_2\rangle_{AB}$.
In particular, set $\beta_1=\frac{1}{4}$ and $\beta_2=\frac{9+\sqrt{13}}{24}$.
One has $C(|\phi_1\rangle_{AB})=C(|\phi_2\rangle_{AB})=\frac{\sqrt{11}}{4}$ and $\widetilde{C}(|\phi_1\rangle_{AB})=\widetilde{C}(|\phi_2\rangle_{AB})=\frac{11}{36}$, which implies that both of the entanglement measures cannot tell the difference between the entanglements
of $|\phi_1\rangle_{AB}$ and $|\phi_2\rangle_{AB}$. However, our ICEM shows that $|\phi_1\rangle_{AB}$ is more entangled than $|\phi_2\rangle_{AB}$, $\mathcal{C}(|\phi_1\rangle_{AB})=0.5139>\mathcal{C}(|\phi_2\rangle_{AB})=0.5126$. In this sense, ICEM has advantages over concurrence $C$ and the concentratable entanglement measure $\widetilde{C}$. By taking another glance at the definitions of these measure, one sees naturally that the ICEM contains more information about the quantum states than the concurrence and the concentratable entanglement measure do, as ICEM takes into account all the coefficients of the characteristic polynomial in Eq. (\ref{eigen}), while the latter ones depend only on the coefficient $a_2$.
\begin{figure}
\centering
    \includegraphics[width=8cm]{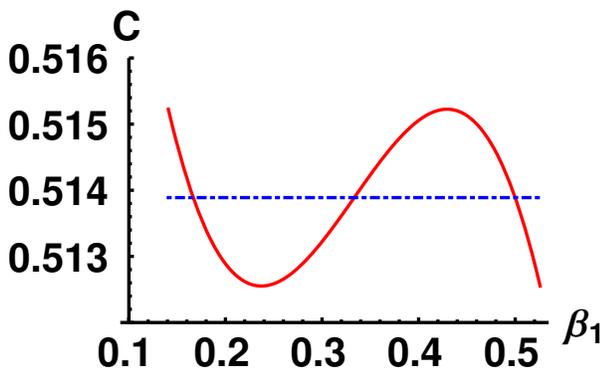}\\
\caption{The blue (dot-dashed) line is the ICEM of $|\phi_1\rangle_{AB}$, and the red (solid-line) line represents the ICEM of $|\phi\rangle$, as a function
of the states given by the points on the ellipse in Fig. 1.}\label{2}
\end{figure}

Moreover, $\mathcal{C}(|\psi\rangle_{AB})$ is related to the state transition under LOCC. ICEM in Eq. (\ref{def}) can be rewritten as $\mathcal{C}(|\psi\rangle_{AB})=\sum_{i=0}^r \mathcal{C}_i(|\psi\rangle_{AB})$, where $\mathcal{C}_i(|\psi\rangle_{AB})=\frac{P_r^i}{2^r}[1-\mathrm{Tr}(\rho^{i+1}_A)]$, $i=0,1,...,r$. We have the following criterion on on state transformation under LOCC, see proof in Supplemental Material IV.

{\it Proposition 1.}  If a state $|\psi\rangle_{AB}$ can be transformed to $|\phi\rangle_{AB}$ under LOCC, then $\mathcal{C}_i(|\psi\rangle_{AB})\geq \mathcal{C}_i(|\phi\rangle_{AB})$ for $i=1,...,r$.

As an example let us consider \cite{fanh} $|\psi\rangle_{AB}=\sqrt{0.5}|00\rangle+\sqrt{0.4}|11\rangle+\sqrt{0.1}|22\rangle$
and $|\phi\rangle_{AB}=\sqrt{0.55}|00\rangle+\sqrt{0.3}|11\rangle+\sqrt{0.15}|22\rangle$.
By calculation we have $\mathcal{C}_1(|\psi\rangle_{AB})=0.29\leq \mathcal{C}_1(|\phi\rangle_{AB})=0.2925$ while $\mathcal{C}_2(|\psi\rangle_{AB})=0.2025\geq \mathcal{C}_2(|\phi\rangle_{AB})=0.2008125$.
From Proposition 1 we have that neither $|\psi\rangle_{AB}\to |\phi\rangle_{AB}$ nor $|\phi\rangle_{AB}\to |\psi\rangle_{AB}$, which is in consistent with the conclusion from
the Nielsen theorem \cite{man}.

{\it ICEM for Multipartite systems}--Now consider general $n$-qudit states. We can define the averaged informationally complete entanglement measure for n-partite pure state $|\psi\rangle_{A_1A_2...A_n}$ as
\begin{eqnarray}\label{def2}
\mathcal{C}(|\psi\rangle_{A_1\cdots A_n})=\sum_{\{A\}}\frac{\mathcal{C}(|\psi\rangle_{A\bar{A}})}{2^{n}-2},
\end{eqnarray}
where $A$ and $\bar{A}$ is a bipartition of the whole n-qudit system ${A_1A_2...A_n}$, $\bar{A}$ is the complement of $A$, the summation goes over all possible bipartitions. When $A= \emptyset$ or $A= A_1A_2...A_n$, we define $\mathcal{C}(|\psi\rangle_{A\bar{A}})=0$.
It is verified that if $\mathcal{{C}}(|\psi\rangle_{A_1\cdots A_n})=0$, then $|\psi\rangle_{A_1,A_2,...,A_n}$ is fully separable.

Instead of arithmetic mean of ICEM defined in Eq. (\ref{def2}), one can also use the geometric mean to give another entanglement measure for multipartite states,
$$
\mathcal{\widetilde{C}}(|\psi\rangle_{A_1\cdots A_n})
=\left(\Pi_{\{A\}}\mathcal{C}(|\psi\rangle_{A\bar{A}})\right)^\frac{1}{2^n-2}.
$$
$\mathcal{\widetilde{C}}(|\psi\rangle_{A_1\cdots A_n})$ can be used to determine whether a pure state is genuine entangled.
$\mathcal{\widetilde{C}}(|\psi\rangle_{A_1\cdots A_n})\neq 0$ means that any possible bipartition is not separable, that is, $|\psi\rangle_{A_1,A_2,...,A_n}$ is a genuine entangled pure state.

As for mixed multipartite states $\rho_{A_1\cdots A_n}$, averaged ICEM is defined by the convex roof extension, $E(\rho_{A_1\cdots A_n})=\min_{\{p_i, |\psi_i\rangle_{A_1\cdots A_n}\} }\sum p_iE(|\psi_i\rangle_{A_1\cdots A_n})$, where $E$ stands for either $\mathcal{C}$ or $\mathcal{\widetilde{C}}$, with the minimization taking over all possible pure state decompositions $\{p_i, |\psi_i\rangle_{A_1\cdots A_n}\}$ of $\rho_{A_1\cdots A_n}$. Similar to the case of multipartite pure states, if $\mathcal{C}(\rho_{A_1\cdots A_n})=0$, then $\rho_{A_1\cdots A_n}$ is fully separable, as each $|\psi_i\rangle_{A_1\cdots A_n}$ of the ensemble $\{p_i, |\psi_i\rangle_{A_1\cdots A_n}\}$ is fully separable. Whereas, $\mathcal{\widetilde{C}}(\rho_{A_1\cdots A_n})\neq 0$ does not imply that $\rho_{A_1\cdots A_n}$ is a genuine entangled mixed state. In fact, $\mathcal{\widetilde{C}}(\rho_{A_1\cdots A_n})\neq 0$ only shows that at least one $|\psi_i\rangle_{A_1\cdots A_n}$ of the ensemble $\{p_i, |\psi_i\rangle_{A_1\cdots A_n}\}$ is a genuine entangled state, but the combination of all $|\psi_i\rangle_{A_1\cdots A_n}$ $(i=1,2,...,n)$ may still be biseparable.

Next we consider the efficient computation of entanglement for multi-qudit pure states.
A fundamental aspect is that any ICEM $\mathcal{C}(|\psi\rangle_{AB})$ of a bipartite state $|\psi\rangle_{AB}$ can be efficiently estimated on a quantum computer.
In Fig. 3 the $k$th ancilla qubit is used to perform a controlled SWAP test on the $k$th and $(k+1)$th copy of $|\psi\rangle_{AB}$.
Given $|\psi\rangle_{AB}$ of a multi-qudit pure state under bipartition $A$ and $B$ with $R(|\psi\rangle_{AB})=r+1$, as well as $r$ qubit states.
Let $p(\vec{z})$ be the probability of measuring the $z$ bitstring on the $r$ control qubits. Denote $\vec{z}\in\{0,1\}^r$ the set of all such bitstrings. We have the following proposition, see proof in the Supplemental Material V.
\begin{figure}
\centering
\includegraphics[width=8cm]{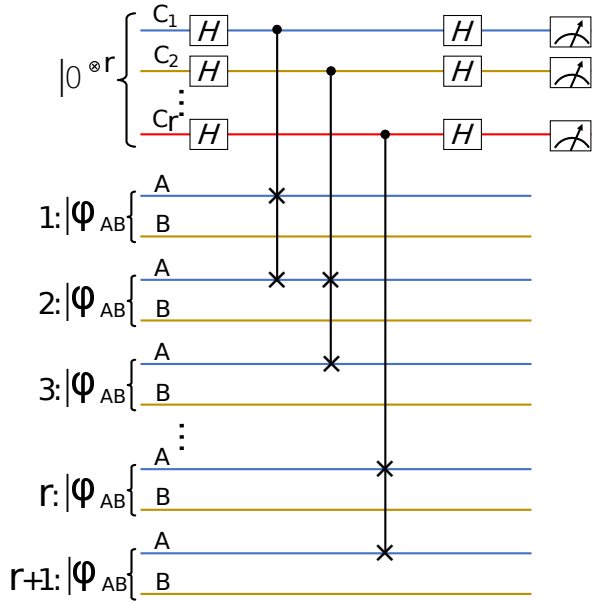}\\
\caption{Circuit for calculating $\mathcal{C}(|\psi\rangle_{AB})$. Given $r+1$ copies of the quantum state $|\psi\rangle_{AB}$ and $r$ ancilla qubits. The $r$ ancilla qubits are applied one by one by SWAP test by employing the $k$th ($k=1,2,...,r$) ancilla to perform a controlled swap test on the $k$th and $(k+1)$th copy of $|\psi\rangle_{AB}$.}\label{2}
\end{figure}

{\it Proposition 2.} The ICEM of a pure quantum state $|\psi\rangle_{AB}$ with $R(|\psi\rangle_{AB})=r+1$ can be computed from the outcomes of the $r$-qubit SWAP test,
\begin{eqnarray}\label{th2}
\mathcal{C}(|\psi\rangle_{AB})=1-p(\vec{0}_r),
\end{eqnarray}
where $p(\vec{0}_r)$ denotes the probability of obtaining the all-zero result from the SWAP test on the $r$-qubit control states.

Proposition 2 shows that $\mathcal{C}(|\psi\rangle_{AB})$ can be computed by performing the SWAP test on subsystem $A$ and adding the probabilities of the measurement outcomes that the controlled qubits are in the state $|0\rangle$, which give rise to the computation of  $\mathcal{{C}}(|\psi\rangle_{A_1\cdots A_n})$ and $\mathcal{\widetilde{C}}(|\psi\rangle_{A_1\cdots A_n})$.

{\it Discussion and summary.}---The conclusion given in Lemma can be also generalized to other quantum correlations besides entanglement, as (\ref{th}) links $\mathcal{T}(\rho)$, $\mathcal{A}(\rho)$ and $\vec{\lambda}(\rho)$ together and establishes one-to-one correspondence among them. As each element in $\mathcal{T}(\rho)$ is non-decreasing, one can built new kinds of quantum correlations via the convex roof construction of $\mathcal{T}(\rho)$ on average under local operations. For example, one may use $\mathcal{T}(\rho)$ instead of $E_k(|\psi\rangle)$ given in \cite{vidal2} (the sum of the $n-k$ smallest $\lambda's$) to investigate the related probability $p(|\psi\rangle\to |\phi\rangle)$ of obtaining state $|\phi\rangle$ from $|\psi\rangle$ under LOCC. One can give a necessary condition for a complete connected quantum networks \cite{lmx2}, for instance, any $\mathcal{\widetilde{C}}(|\psi\rangle_{A\bar{A}})=0$ implies that $|\psi\rangle_{A_1\cdots A_n}$ is not an $n$-partite complete connected quantum network. The ICEMs can also be used to establish genuine entanglement witnesses. For example, as $\mathcal{C}_1(|\psi\rangle_{AB})=\frac{1}{2}[1-\mathrm{Tr}(\rho^{2}_A)]$, one gets $\mathcal{C}_1(|\psi\rangle_{A_1A_2|B})\leq \mathcal{C}_1(|\psi\rangle_{A_1|A_2B})+\mathcal{C}_1(|\psi\rangle_{A_2|A_1B})$. Since $1+\mathrm{Tr}(\rho^2_{AB})\geq \mathrm{Tr}(\rho^2_{A})+\mathrm{Tr}(\rho^2_{B})$ \cite{gy}, one obtains an genuine entanglement witness by using a similar method presented in \cite{xsb}.

In summary, we have introduced a method of providing general framework to quantify the quantum entanglement in an informationally complete way. The entanglement measures called ICEMs have been presented to characterize and quantify the multipartite entanglement of arbitrary $n$-qudit pure states. By specific examples we have shown the advantages of ICEMs in distinguishing the entanglement of different states whose entanglement can not be identified by some well-know entanglement measures. We have also shown that the entanglement in and between the subsystems of a composite quantum state can be efficiently estimated on a quantum computer by implementing one-by-one SWAP tests. Moreover, the ICEMs are also of paramount importance to state transition under LOCC. Furthermore, the arithmetic and geometric means of ICEMs can be used to detect the full separability, entanglement or genuine multipartite entanglement faithfully on quantum devices.

\bigskip
\noindent{\bf Acknowledgments}\, \,
This work was supported in part by the National Natural Science Foundation of China (NSFC) under Grants 11847209; 12075159; 12171044; 11975236 and 11635009; Beijing Natural Science Foundation (Grant No. Z190005); Academy for Multidisciplinary Studies, Capital Normal University; Shenzhen Institute for Quantum Science and Engineering, Southern University of Science and Technology (No. SIQSE202001); Academician Innovation Platform of Hainan Province; China Postdoctoral Science Foundation funded project No. 2019M650811 and  China Scholarship Council No. 201904910005.

\end{document}